\documentclass[prl,letterpaper,twocolumn,showpacs]{revtex4}  
\usepackage{times,xspace}  
\usepackage{amsbsy,amssymb,amsmath,bm}    
\usepackage{graphicx,color,epsfig,rotate}  
\usepackage{fancyhdr}  
  
\def\bbbc{{\mathchoice {\setbox0=\hbox{$\displaystyle\rm C$}\hbox{\hbox  
to0pt{\kern0.4\wd0\vrule height0.9\ht0\hss}\box0}}  
{\setbox0=\hbox{$\textstyle\rm C$}\hbox{\hbox  
to0pt{\kern0.4\wd0\vrule height0.9\ht0\hss}\box0}}  
{\setbox0=\hbox{$\scriptstyle\rm C$}\hbox{\hbox  
to0pt{\kern0.4\wd0\vrule height0.9\ht0\hss}\box0}}  
{\setbox0=\hbox{$\scriptscriptstyle\rm C$}\hbox{\hbox  
to0pt{\kern0.4\wd0\vrule height0.9\ht0\hss}\box0}}}}

\newcommand{\ignore}[1]{}  
\newcommand{\mComment}[1]{}  
\newcommand{\gComment}[1]{}  
\newcommand{\jComment}[1]{}  
\newcommand{\rComment}[1]{}  
\newcommand{\lComment}[1]{}  
  
\begin{document}  
\title{Spin Supersolid in Anisotropic Spin-One Heisenberg Chain}  
\author{P. Sengupta$^{1,2}$ and C. D. Batista$^1$ }  
\affiliation{$^1$Theoretical Division, Los Alamos National Laboratory, Los Alamos, NM 87545 \\  
$^2$ MST-NHMFL, Los Alamos National Laboratory, Los Alamos, NM 87545}  
  
\date{\today}

\begin{abstract}  
We consider an $S=1$ Heisenberg chain with strong exchange ($\Delta$) and 
single--ion uniaxial anisotropy ($D$) in  a magnetic field ($B$) along the symmetry axis. 
The low energy spectrum is described by  
an effective $S=1/2$ XXZ model that acts on two different low energy  
sectors for a given window of fields. The vacuum of each sector exhibits Ising-like antiferromagnetic  
ordering that coexists with the finite spin stiffness obtained  
from the exact solution of the effective XXZ model. In this way, we demonstrate  
the existence of a spin supersolid phase. We also compute the full $\Delta-B$ quantum phase diagram 
by means of a quantum Monte Carlo simulation.  
\end{abstract}  
  
  
\maketitle %
\thispagestyle{fancy}

  
One of the primary goals in the study of strongly correlated systems is  
the search for novel states of matter. These novel states usually emerge from  
competing interactions, each of which tends to stabilize distinct  
orderings. Different outcomes can result from such competition:  
a) none of the competing phases prevail, b) one interaction becomes  
dominant at low energies and stabilizes the corresponding ordering, c)   
more than one competing orderings coexist in a new phase. In the last  
case, the coexistence can be homogeneous or inhomogeneous. The supersolid  
phase is one of the simplest examples of homogeneous coexistence of two  
different phases (solid and superfluid) in bosonic systems. However, this novel state  
has been elusive both from an experimental \cite{Kim04} and a theoretical   
\cite{SStheory} viewpoint. While it is still unclear whether a supersolid (SS)  
state can be stabilized in the continuum, there are several numerical studies  
which show that a SS phase can be stabilized in the presence of a periodic   
potential or underlying lattice \cite{Triangular,Sengupta05,Batrouni06}. Surprisingly,   
this numerical effort has not been complemented whatsoever with rigorous analytical  
treatments that can demonstrate the existence of the SS phase, at least  
in certain limits.   
  
The SS state is easier to stabilize on a lattice  because the    
lattice parameter of the ``solid phase" or charge density wave cannot relax to any   
arbitrary value (it has to be an integer multiple of the underlying lattice parameter).   
There are two natural realizations of bosonic gases on a lattice: atomic gases in  
optical lattices or periodic substrates and quantum magnets (spin lattices).   
Although most of the previous numerical work has been focused on the former systems   
\cite{Triangular,Sengupta05,Batrouni06}, we have shown  
recently \cite{Sengupta07} that a class of spin Hamiltonians, which describe real magnets  
to a very good approximation, also contain a SS phase in their quantum phase  
diagram. In particular, the models for hard core bosons on frustrated lattices  
that have been considered in the context of atomic gases can also be generated as  
low energy effective theories of frustrated spin--dimer systems \cite{Sengupta07}.  
The interest and the number of numerical works on spin SS phases is rapidly  
increasing \cite{Ng06,Laflorencie07,Schmidt07}.  
  
The area of 1D magnetism has also attracted a lot of attention during the   
last few decades. This interest was mainly triggered by the successful application   
of field theory techniques for finding relevant and solvable low energy effective theories.  
A remarkable accomplishment was achieved by Haldane \cite{Haldane83},   
who pointed out a qualitative difference between the low energy spectra of integer   
and half--odd--integer Heisenberg chains. The gapped Haldane phase has peculiar properties with  
measurable consequences, like the $S=1/2$ end--chain states of open $S=1$ Heisenberg  
chains \cite{Batista98}.

In spite of the intense effort devoted to the study of 1D Heisenberg like spin models,   
the search for new states and behaviors is far from being complete.  
In this Letter, we demonstrate that the low energy spectrum of an S=1 Heisenberg  
chain with uniaxial exchange and single--ion anisotropy consists of two sectors, each  
of which can be mapped into the exactly solvable (S=1/2) XXZ model for the limit of  
strong anisotropy. By exploiting this mapping, we also demonstrate that the ground state  
is a SS (Ising--like spin ordering coexisting with a finite spin stiffness)   
in a finite range of magnetic field. This is a surprising result considering that 1D 
solid phases are usually unstable when the particle density (magnetization for spin
systems) deviates from the corresponding 
commensurate value. This occurs because each added (or removed) particle introduces 
a soliton in the ground state, something that is particularly clear in the  
bosonization treatments \cite{Giamarchi04}. In our case, 
the solid (Ising) phase stabilized by the exchange anisotropy and  the solitons 
remain as massive excitations for a finite range of densities
due to the presence of a significant single ion anisotropy. This exceptional 
behavior leads to the stabilization of the supersolid state.
In addition, we compute the   
full quantum phase diagram of $H$ that covers different regimes  
of anisotropy and was computed using a quantum Monte Carlo (QMC) algorithm.

We start by considering an $S=1$ Heisenberg model with uniaxial exchange  
and single--ion anisotropies on a linear chain ${\cal L}$ of $L$ sites:  
\begin{equation}  
H= \sum_{i=1,L} J_{\perp} (S^x_i S^x_{i+1} + S^y_i S^y_{i+1})  
+ J_z S^z_i S^z_{i+1} +  D {S^z_i }^2 - B S^z_i.   
\label{eq:s1hamil}  
\end{equation}  
Here $J_z>0$, $D>0$ and we define $L+1 \equiv 1$ to impose periodic boundary conditions (PBC).  
Since $H$ is invariant under  
spin--rotations along the $z$--axis, the $z$-component of the magnetization,   
$M=\sum_i  S^z_i $, is a good quantum number. We will assume that   
$\Delta = J_z/J_{\perp} \gg 1$, $J_{\perp} \ll D$, $J_{\perp} \ll J_z-D$ and $J_z > D$. Therefore, we can treat $J_{\perp}$  
as a perturbation. For $J_{\perp}=0$, $H$ becomes diagonal in the basis of   
eigenstates of the set of operators $\{S^z_i\}$. The  
chain system consists of two interpenetrating sublattices ${\cal A}$ and ${\cal B}$. Since   
$J_z > D >0$, there are only two ground states for $B=0$ (i.e., $M=0$): $S^z_i=\pm1$  
for $i \in {\cal A}$ and $S^z_i=\mp1$ for $i \in {\cal B}$:  
\begin{eqnarray}  
|\psi^0_{\cal A}\rangle &=& \frac{1}{2^{L/2}} \prod_{i\in {\cal A}} S^{+}_i \prod_{i\in {\cal B}}  
S^{-}_i |00.....0\rangle  
\nonumber \\   
|\psi^0_{\cal B}\rangle&=&  
\frac{1}{2^{L/2}} \prod_{i\in {\cal B}} S^{+}_i \prod_{i\in {\cal A}}  
S^{-}_i |00.....0\rangle  
\label{psi}  
\end{eqnarray}  
where $|00.....0\rangle$ is the reference state in which all the spins are in the   
eigenstate of $S^{z}_i$ with eigenvalue zero: $S^{z}_i|00.....0\rangle=0$ $\forall i \in {\cal L}$.  
Any other state with $M=0$  
has an energy $2J_z-D$ or higher. Let us consider now the case  
$0<M<L/2$. In this case, the low energy subspace can again be divided into  
two different sectors or subspaces ${\cal S^A}$ and ${\cal S^B}$. The  
subspace ${\cal S}^{A}$ is generated by the following basis of states:  
\begin{equation}  
\{ | \phi^{\cal A}_{i_1 ...i_M} \rangle= \frac{1}{2^{M/2}} S^{+}_{i_1} S^{+}_{i_2}....S^{+}_{i_M} |\psi_{\cal A}\rangle \}  
\label{sa}  
\end{equation}  
where the sites $\{ i_1 ...i_M \} \in {\cal B}$ are all different: $i_1<i_2 ...<i_M$ . In the same way,   
the subspace ${\cal S^B}$ is generated by the basis of states:  
\begin{equation}  
\{ |\phi^{\cal B}_{j_1 ...j_M} \rangle= \frac{1}{2^{M/2}} S^{+}_{j_1} S^{+}_{j_2}....S^{+}_{j_M} |\psi_{\cal B}\rangle \}  
\end{equation}  
where the sites $\{ j_1 ...j_M \} \in {\cal A}$ are all different: $j_1<j_2 ...<j_M$ . The energy of any other   
state that has magnetization $M$ and is orthogonal to  ${\cal S}={\cal S^A} \oplus {\cal S^B}$ is higher   
by at least $2J_z-D$. Therefore, we can use degenerate perturbation theory to solve the low energy spectrum  
for small $J_{\perp}$: $J_{\perp} \ll 2J_z-D$ and $J_{\perp} \ll D$. The subspaces ${\cal S^A}$  
and ${\cal S^B}$ cannot be connected by any finite order process in the thermodynamic limit $L \to \infty$.  
Consequently, we have two identical and disconnected low energy theories on each sector. From now on,  
we will consider one of them  without loss of generality.    
The low energy subspace ${\cal S^A}$ (Eq.\ref{sa}) can be mapped into the Hilbert space for $M$ hard core bosons   
on the $L/2$ sites of the ${\cal B}$ sublattice:  
\begin{equation}  
| \phi^{\cal A}_{i_1 ...i_M} \rangle= b^{\dagger}_{i_1}  b^{\dagger}_{i_2} .... b^{\dagger}_{i_N} |0_{\cal A}\rangle,  
\end{equation}   
where $i_1 <i_2 ...i_N \in {\cal B}$.    
The relation between the hard core bosons and the original spin operators is given by:  
\begin{equation}  
S^{+}_i= \sqrt{2} b^{\dagger}_i,  \;\;\;  S^{-}_i= \sqrt{2} b^{\;}_i,  \;\;\; S^z_{i}=-1+n_i,  
\end{equation}  
where $i\in {\cal B}$ and $n_i=b^{\dagger}_{i} b^{\;}_{i}$.   
We emphasize that these relations are  
only valid within the low energy subspace ${\cal S^A}$.  
In addition, $S^z_{i}=1$ for $i\in {\cal A}$, which implies  
that the two--spin correlators $\langle S^z_i S^z_j \rangle$ and $\langle S^+_i S^-_j \rangle$ have   
the following expressions up to quadratic corrections in the perturbative parameter $J_{\perp}/J_z$:  
\begin{eqnarray}  
\langle S^z_j S^z_{j+r} \rangle &=& e^{i \pi r} (1-\langle n_j\rangle-\langle n_{j+r}\rangle)+ \langle n_j n_{j+r} \rangle  
\label{szsz}  
\\  
\langle S^+_j S^-_{j+r} \rangle &=& 2 \langle b^{\dagger}_j b^{\;}_{j+r} \rangle  
\label{s+s-}  
\end{eqnarray}  
where $b^{\dagger}_i \equiv 0$ and $n_i \equiv 0 \; \forall i\in {\cal A}$.  
After doing a canonical transformation and projecting  
out the high energy states:  
\begin{equation}  
{\tilde H} = P e^{-S} H e^s P = {\tilde H}_{\cal A} + {\tilde H}_{\cal B}  
\end{equation}  
we obtain the following expression for the low energy effective model, ${\tilde H}_{\cal A}$,   
that acts on the sector ${\cal S^A}$:  
\begin{equation}  
{\tilde H}_{\cal A} = LC + \sum_{i\in {\cal B}} t (b^{\dagger}_{i} b^{\;}_{i+2} + b^{\dagger}_{i+2} b^{\;}_{i})   
- \mu n_i + V  n_i n_{i+2}  
\end{equation}  
where $L+2\equiv2$ (PBC) and  
\begin{eqnarray}  
C &=& -\frac{ J_{\perp}^2}{3J_z -2D},  
\;\;\;\;  
t = - \frac{J_{\perp}^2}{2(J_z-D)},  
\nonumber \\  
\mu &=& \frac{J_{\perp}^2}{J_z} -2t + 4C + B + D - 2J_z,  
\nonumber \\  
V &=&  - \frac{J_{\perp}^2}{J_z}-2t +2C.  
\end{eqnarray}    
We note that $|V|<-2t$. ${\tilde H}_{\cal A}$ is the so--called $t-V$ model (or S=1/2 XXZ Hamiltonian \cite{Matsubara56}) and is   
exactly solvable by the Bethe Ansatz method \cite{Baxter}. The ground state is a Luttinger liquid for $|V|<|2t|$ and $-2|t|<\mu<2V+2|t|$.   
The asymptotic behavior of the transverse and longitudinal two point correlators can be    
obtained with the bosonization method \cite{Giamarchi04}:  
\begin{eqnarray}  
\langle n_j n_{j+r} \rangle &=& {\tilde \rho}^2 - \frac{K}{2\pi^2r^2} + C_1 r^{-2K} \cos{(2 \pi {\tilde \rho} r)}    
\label{nn} \\  
\langle b^{\dagger}_j b^{\;}_{j+r}  \rangle &=& C_2 r^{-g(K)} \cos{\alpha r}   
+ C_3  r^{-1/2K} \cos{\pi r}  
\end{eqnarray}  
where $g(K)=2K+1/(2K)$, $\alpha=[(2{\tilde \rho}-1)\pi$ , $K$ is the Luttinger liquid (LL) parameter   
and ${\tilde \rho}=\langle n_j\rangle \; \forall i\in {\cal B}$.   
A schematic contour map of $K$ as a function  
of $V/t$ and ${\tilde \rho}$ can be found in Ref.\cite{Giamarchi04}. The wave--length, $\lambda$, of the oscillations of the   
density--density correlator is the mean separation between bosons, i.e., $\lambda=1/{\tilde \rho}$.   
Therefore, the longitudinal spin--spin  
correlator has two oscillatory components. The first component (first term of Eq.\ref{szsz})   
has a constant amplitude and the wave--length is equal to two lattice parameters. This contribution comes from   
the ``solid'' or Ising component of the SS phase. The second contribution (second term of Eq.\ref{szsz}) comes  
from the LL component and consequently decays with a power law according to Eq.\ref{nn}. The density of the   
LL in the supersolid phase ($\rho\equiv{\tilde \rho}/2$ since ${\tilde \rho}$ the density on the ${\cal B}$   
sublattice) can be extracted from the wave--length, $2\lambda$, of this oscillatory component.   
  
\begin{figure}[!htb]  
\hspace*{-1.3cm}  
\includegraphics[angle=0,width=18.0cm]{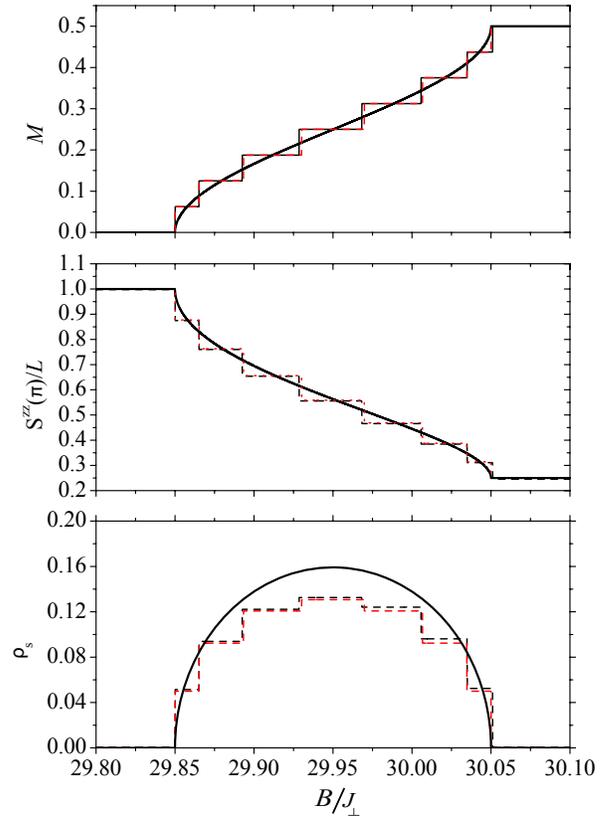}  
\vspace{-2cm}  
\caption{ (Color online) Magnetization along the field direction $M_z$ (a),   
longitudinal structure factor $S^{zz}(\pi)/L$ (b), and   
spin stiffness $\rho_s$ (c) as a function of field for $J_z=20J_{\perp}$   
and $D=10J_{\perp}$. The dotted (dashed) line corresponds to the exact   
solution of the original (effective) model in a 16 sites chain. The full   
line corresponds to the exact solution of the effective model in the  
thermodynamic limit $L \to \infty$.}  
\label{dimer}  
\end{figure}

In order to test the accuracy of our effective model, we used the LANCZOS method to compute the exact ground state   
of the original Hamiltonian, $H$, in a $L=16$ sites chain. Fig.\ref{dimer}a shows a comparison between the magnetization  
as function of field, $M(B)$, obtained with the original and the effective models for $J_z=20J_{\perp}$ and $D=10J_{\perp}$.  
The curves are practically indistinguishable. The full line is the $M(B)$ curve obtained with ${\tilde H}_{\cal A}$   
in the thermodynamic limit $L\to \infty$. Fig.\ref{dimer}b shows a similar comparison for the field dependence of the   
longitudinal spin structure factor:  
\begin{equation}   
S^{zz}(q)={1\over L}\sum_{j,k}e^{-iq(j-k)}\\    
\langle  S^z_jS^z_k\rangle.    
\end{equation}  
Again, practically the same curves are obtained with $H$ and ${\tilde H}_{\cal A}$ for a 16 sites chain. The finite  
value of $S^{zz}(q)/L$ for $L\to \infty$ indicates the existence of Ising ordering. Finally, Fig.\ref{dimer}c shows the same  
comparison for the field dependence of the spin stiffness $\rho_s$ or superfluid density in the bosonic language.   
To define $\rho_s$, we will use units of $\hbar=1$, lattice parameter $a=1$  and effective mass of the bosons $m=1/8t=1$.  
$\rho_s$  can be defined as the response to the gauge field generated by an infinitesimal flux $\phi$ threading   
the ring (chain with PBC),  
$\rho_s= \frac{\partial^2\epsilon}{\partial^2\phi}$,  
where $\epsilon$ is the free energy per site (ground state energy density at $T=0$).   
According to Figs.\ref{dimer}b and \ref{dimer}c, the SS phase appears for $B_{c1}<B<B_{c2}$,  
where  
\begin{eqnarray}  
B_{c1}&=& 4t+2J_z-D-4C-\frac{J_{\perp}^2}{J_z}  
\nonumber \\  
B_{c2}&=& 2V+2J_z-D-4C-\frac{J_{\perp}^2}{J_z}  
\label{bcs}  
\end{eqnarray}  
are determined by the conditions $\mu_{c1}=2t$ and $\mu_{c2}=2V-2t$. Two different   
Ising phases appear on both sides: IS1 for $B<B_{c1}$ and IS2 for $B_{c2}<B$.  
In the dilute limit, $B\to B_{c1}^+$ (see Fig. \ref{dimer}c), the superfluid density approaches the value of   
the total density $\rho$ at $T=0$. This is the known result for bosons in the continuum that  
is recovered in the dilute limit of the lattice system ($\rho a \to 0$).  

\begin{figure}[!htb]  
\hspace*{-0.5cm}  
\includegraphics[angle=-90,width=9.0cm]{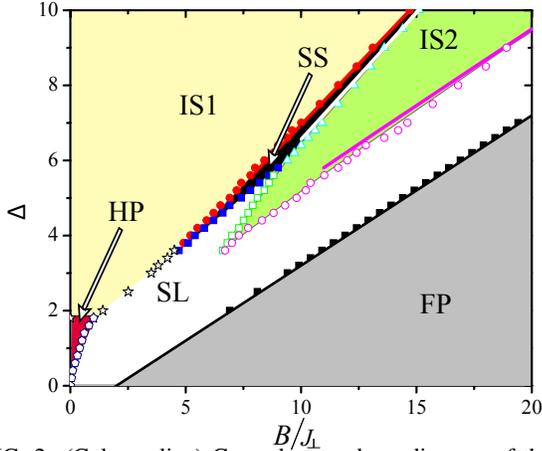}  
\vspace{-0.7cm}  
\caption{(Color online) Ground state phase diagram of the $S=1$ Heisenberg model in one  
dimension in the $\Delta-B$ parameter space with $D=\Delta/2$. The spin-gapped  
Ising-like phases IS1 ($m_z=0$) and IS2 ($m_z=0.5$) phases have long range   
diagonal order ($S^{zz}/L > 0$ for $L\to \infty$) whereas the SL phase has a finite stiffness  
($\rho_s>0$). The spin SS has simultaneous  
long range diagonal order and finite stiffness while both quantities are zero in the Haldane  
phase (HP). The symbols are   
results obtained from QMC simulations. The solid lines are  
the phase boundaries, $B_{ci}$ with $1\leq i\leq 4$ (see Eqs.\ref{bcs} and \ref{bethe})   
obtained from the effective low--energy models valid for $\Delta \gg 1$.}  
\label{fig:phasediagram}  
\end{figure}

The Ising order IS2 corresponds to  
the state in which $S^z_i=1 \; \forall i\in {\cal A}$ and $S^z_i=0 \;\forall i\in {\cal B}$. If we keep increasing the field,  
ground states with higher values of $M$ will be eventually stabilized. For $L/2 \leq M <L$, we can again identify a low   
energy subspace generated by the states:   
$\{ |\psi_{i_1 ...i_M}\rangle = \frac{1}{2^{M/2}} S^{+}_{i_1} S^{+}_{i_2}....S^{+}_{i_M}|00...0\rangle \}$ where $i_n$ denotes  
now {\it any} site of ${\cal L}$ and $i_1<i_2 ...<i_M$. Again this subspace can be mapped into the one for hard--core  
bosons on the {\it full} chain ${\cal L}$ and the effective low--energy Hamiltonian is again a $t-V$ model with  
$t=J_{\perp}$, $V=J_z$ and $\mu=B-D$. Since $V\gg t$, the IS2 phase ($M=L/2$) remains stable over a broad region of  
field. However, as shown in Fig.\ref{fig:phasediagram}, a second order transition to a spin liquid (SL) phase   
occurs at the critical value $B=B_{c3}$ that  
is obtained from the exact solution of the Bethe-Ansatz equations for the $t-V$ model \cite{Gaudin}:  
\begin{equation}  
B_{c3}= D + J_{\perp}\Delta+\frac{4\pi J_{\perp}\sinh{\gamma}}{\gamma} \sum_{n=0}^{\infty}   
\frac{1}{\cosh{[(2n+1)\pi^2/2\gamma]}}  
\label{bethe}  
\end{equation}  
where $\Delta=2\cosh{\gamma}$. The three transitions at $B_{c1}$, $B_{c2}$  
and $B_{c3}$ belong to the Dzhaparidze--Nersesyan--Pokrovsky--Talapov   
universality class \cite{Dzhaparidze}. Finally, the system becomes fully saturated  
at the critical field $B_{c4}=2J_z+2J_{\perp}+D$ (this expression is also valid away from the strongly   
anisotropic limit $\Delta \gg 1$).

In order to extend the quantum phase diagram of $H$ away from the  
strongly anisotropic limit, we have used the Stochastic Series expansion (SSE) \cite{Sandvik99}   
quantum Monte Carlo (QMC) method. The simulations were done on  
finite chains of length $16\leq L \leq 64$. The SSE is a   
finite-temperature QMC based on the Taylor expansion of the partition function,  
$e^{-\beta H}$. Ground state estimates for the observables are obtained by  
choosing sufficiently large values of the inverse temperature $\beta$. For the   
parameters explored in this study, $\beta=2L$ was found to be sufficient for the  
observables to have converged to their ground state values. The  
so--called Haldane state is the only new phase  
that appears in the full quantum phase diagram (see  
Fig.\ref{fig:phasediagram}) relative the strongly anisotropic limit.    
To characterize the different emergent phases, we computed $M$, $\rho_s$ and $S^{zz}(q)$.   
The spin stiffness, $\rho_s$, is simply obtained by computing the winding number ($W$)   
fluctuations of the world lines:$\rho_s=\langle W^2 \rangle/\beta$.\cite{Pollock87}   
  
Both Ising phases, IS1 and IS2, are marked by a finite value   
of $S^{zz}(Q) \propto L $  and a vanishing value of $\rho_s$ in the limit $L \to \infty$.    
The spin SS phase is characterized by a finite value of both   
$S^{zz}(Q)/L $ {\em and} $\rho_s$ in the same limit, while only   
$\rho_s$ remains finite ($S^{zz}(Q)/L \to 0$) for the SL phase.   
Finally, the Haldane phase is characterized by a hidden  ordering  
\cite{Nijs89} and both quantities, $\rho_s$ and $S^{zz}(Q)/L$ go to zero in the thermodynamic  
limit. Since all these quantities are finite for finite size systems and estimates for   
$L \to \infty$ are obtained from finite-size scaling.   
  
The results of the simulation are summarized in the ground state phase diagram   
for the $\Delta-B$ parameter space with $D=\Delta/2$ (Fig.\ref{fig:phasediagram}).   
For $\Delta \leq 1$, the ground state is a SL with finite $\rho_s$ above  
a critical field $B_{c0}$ and up to the saturation field $B_{c4}$. Below $B_{c0}$, the   
ground state is in the Haldane phase \cite{Haldane83}. If we increase  
$\Delta$, we find that the Haldane phase (HP)  
is separated from the IS1 phase by a line of critical points. This line  
appears because the excitations that becomes gapless at  
$\Delta=\Delta_c\simeq 1.8$ \cite{Chen03} and $B=0$ have $M=0$ (this is a second order transition between two  
$M=0$ ground states) while the $|M|>1$ excitations remain gapped.  
The Ising--like phase IS1 evolves continuously into the state $|\psi^0_{\cal A}\rangle$ (or $|\psi^0_{\cal B}\rangle$)   
for $\Delta \gg 1$ (see Eq.\ref{psi}).  
This corresponds to the empty band of the effective model ${\tilde H}_{\cal A}$. For $\Delta \lesssim 3.6$,   
increasing the field $B$ induces a transition from the IS1 phase to the SL and from the SL to the  
fully polarized (FP) phase. At larger $\Delta(\gtrsim 3.6)$, the Ising ordering persists along with the SL for a finite range  
of $B$ giving rise to a spin SS that is continuously connected with the SS  
phase that was analytically obtained in the limit $\Delta \gg 1$. For   
$\Delta \gtrsim 5.8$, the SS phase ends up in the IS2 phase, as obtained  
for $\Delta\gg 1$, while the SL phase appears in between for $3.6 \lesssim \Delta \lesssim 5.8$.  
At even higher fields, there is an IS2-SL transition at $B=B_{c3}$ and the system reaches saturation   
($M/L=1$) for $B=B_{c4}$. Both phases and transitions are continuously connected  
with the ones obtained for $\Delta \gg 1$. The phase boundaries, $B_{ci}$ with $1\leq i\leq 4$, obtained from the  
low energy effective models are shown in Fig.\ref{fig:phasediagram} with solid lines. The close agreement with numerics  
for $\Delta \gg 1$ confirms the validity of  the low energy models in this limit.

In summary, we have demonstrated the existence of a spin SS phase induced by field in a one-dimensional  
Heisenberg model with strong uniaxial anisotropy. This demonstration can  
be easily extended to higher dimensions \cite{Sengupta07}. We have also computed the quantum  
phase diagram as a function of the exchange anisotropy $\Delta$ and field $B$.  
In particular, our ground state solution becomes asymptotically exact in the   
strongly anisotropic limit allowing for a full characterization of the SS spin phase.   
This is a remarkable result considering that SS phases found in other models were always obtained  
from numerical or approximated treatments.

We thank B. S. Shastry and T. Giamarchi for helpful discussions.  
LANL is supported by US DOE under Contract No. W-7405-ENG-36.

\end{document}